\documentclass[aps,prb,twocolumn,groupedaddress,showpacs,floatfix]{revtex4}
\usepackage{graphicx}

\begin{document}


\title{Nuclear Spin Coherence in a Quantum Wire}


\author{A. C\'orcoles}
\email[Electronic address: ]{adc48@cam.ac.uk}

\author{C. J. B. Ford}
\author{M. Pepper}
\author{G. A. C. Jones}
\author{H. E. Beere}
\author{D. A. Ritchie}
\affiliation{Cavendish Laboratory, J. J. Thomson Avenue, Cambridge, CB3 0HE, United Kingdom}

\date{\today}

\begin{abstract}
We have observed millisecond-long coherent evolution of nuclear spins in a quantum wire at 1.2 K. Local, all-electrical manipulation of nuclear spins is achieved by dynamic nuclear polarization in the breakdown regime of the Integer Quantum Hall Effect combined with pulsed Nuclear Magnetic Resonance. The excitation thresholds for the breakdown are significantly smaller than what would be expected for our sample and the direction of the nuclear polarization can be controlled by the voltage bias. As a four-level spin system, the device is equivalent to two qubits.
\end{abstract}

\pacs{76.70.Fz, 73.43.Fj, 73.63.Nm, 76.60.Es}

\maketitle

\section{INTRODUCTION \label{s1}}

Nuclear Magnetic Resonance (NMR) in solid-state physics has drawn much attention over the last decade as a candidate for quantum information processing.\cite{DiVincenzo, KaneNature, LeuenbergerLoss, TaylorMarcus, YusaNature} Due to their weak environmental coupling, nuclear spins have very long coherence times. NMR measurements of nuclear spins, however, are challenging unless bulk samples are used, which results in inhomogeneous broadening of the NMR signal. In semiconductor devices, where thermal polarization of nuclei is weak, NMR signals can be significantly enhanced by Dynamic Nuclear Polarization (DNP), a mechanism by which an electron transition between two spin-resolved energy states, or ``spin-flip'', is mediated by a reverse ``spin-flop'' in the nuclear system. Energy conservation makes this process difficult because the energy associated with the orientation of an electron is three orders of magnitude larger than that of a nucleus. This problem can be circumvented by employing different techniques, such as microwave-induced DNP,\cite{Deimling, Dirksen} optical pumping,\cite{Barrett, Kikkawa, SanadaPRL, Kondo} inter-edge-channel scattering in the Quantum Hall regime,\cite{Machida, Dixon} Electron Spin Resonance,\cite{Bowers, Olshanetsky} Quantum Hall Ferromagnetism\cite{Smet,YusaNature} or the breakdown of the integer Quantum Hall Effect (QHE).\cite{Kawamura, Takahashi} As recently demonstrated,\cite{YusaNature} multiple quantum coherence in a nanometer-scale device can be achieved by DNP. In addition, coherence times of tens of seconds have been observed in bulk Si crystals at room temperature, although thermal equilibration times of the order of hours make averaging over several acquisitions impractical.\cite{LaddPRB}

In this paper, we describe a technique for all-electrical coherent control of nuclear spins in a nanometre-scale based on the breakdown of the QHE in a quantum wire (QW). This technique presents a number of advantadges over previous experiments: quantum coherence survives up to 1.2 K for times of the order of a millisecond with a relatively simple sample design and without extreme demands on sample quality. We describe our sample in Sec. \ref{s2}. Our results, including dynamic nuclear polarization, thresholds for the breakdown of the integer QHE and coherent evolution of nuclear spins under pulsed NMR, are presented and discussed in Sec. \ref{s3}.

\begin{figure}[h!]
\includegraphics[width=\columnwidth]{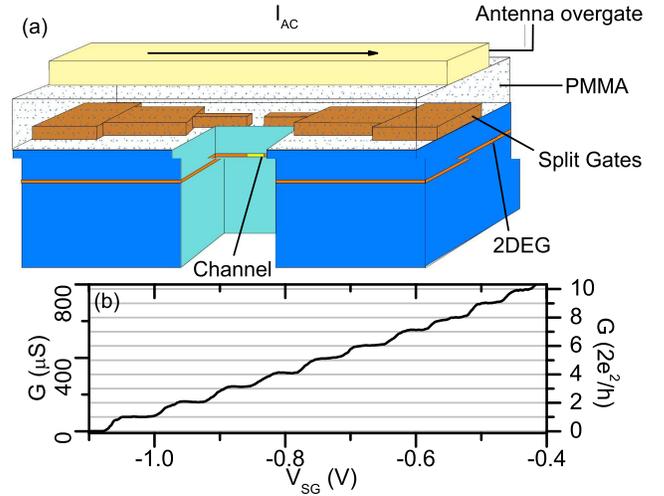}
\caption{(color online). (a) Schematic view of the device (not to scale). Two split gates define the one-dimensional channel. A layer of PMMA isolates the split gates from the antenna overgate used to apply the RF magnetic field; (b) One-dimensional subbands as a function of gate voltage $V_{\rm{SG}}$ at 50 mK and zero magnetic field. The conductance is shown in $\mu$S (left hand vertical axis) and in units of $2\rm{e}^{2}/h$ (right hand vertical axis) \label{f1}}
\end{figure}

\begin{figure*}[t]
\includegraphics[width=2\columnwidth]{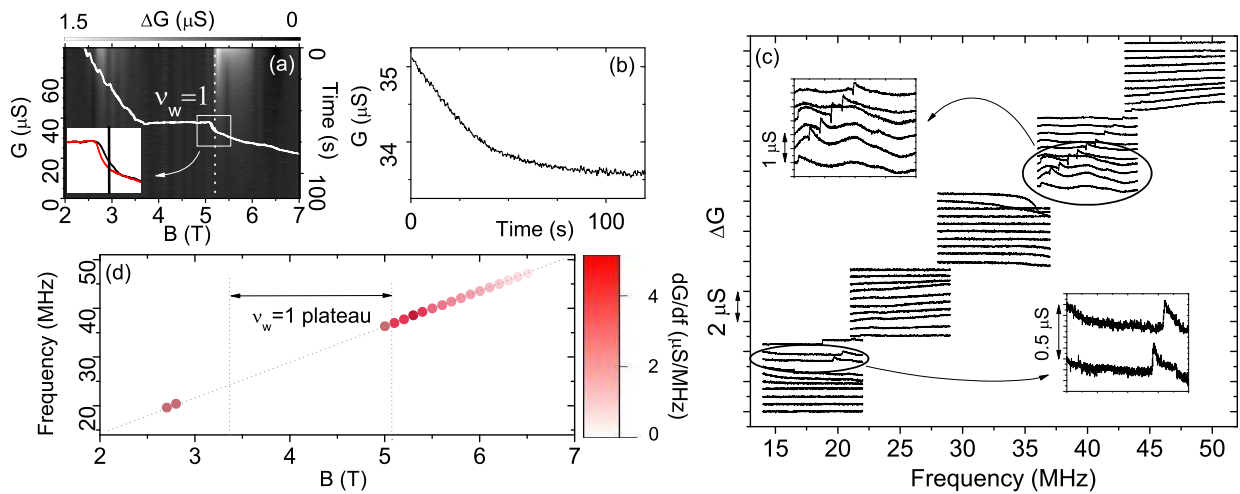}
\caption{(color online). (a) Hysteresis curve for $V_{\rm{SG}}=-0.8$ V (light traces, left hand axis) and conductance-relaxation measurements (grayscale plot, right hand axis) as a function of magnetic field. The inset shows a blow-up of the hysteresis region just below $\nu_{\rm{w}}=1$; (b) Conductance relaxation at $V_{\rm{SG}}=-0.8$ V and $B=5.2$ T [dotted line in (a)]; (c) $^{75}$As NMR signals for $V_{\rm{SG}}=-0.8$ V. Magnetic field varies in steps of 0.1 T from 2 (lowermost curve) to 6.9 (uppermost curve) T. Data have been shifted vertically for clarity. Insets show blow-ups of the regions just below 3 and just above 5 T, where the NMR signals can be seen; (d) Derivative of the NMR signals shown in (c) as a function of magnetic field. \label{f2}}
\end{figure*}

\section{SAMPLE \label{s2}}

Our sample [depicted in Fig. \ref{f1}(a)] consists of a pair of split gates (0.7 $\mu$m wide and 0.7 $\mu$m long) defined on a GaAs/AlGaAs heterostructure. The two-dimensional electron gas (2DEG) is formed 90 nm below the surface, has a density of 1.55x$10^{15}$ m$^{-2}$ and a mobility of 146 m$^{2}$/V s. Separated by a 250 nm layer of crosslinked polymethylmetachrylate (PMMA), a 2.5 $\mu$m wide antenna overgate is deposited on top of the split gates for RF irradiation. We operated our device in the quantum Hall regime, where the one-dimensional energy subbands \cite{vanWees, Wharam} [Figure \ref{f1}(b)] mix with the Landau levels, giving rise to the formation of hybrid magnetoelectric subbands. Transport measurements were performed by standard lock-in techniques in a dilution refrigerator with a base temperature of 50 mK. In order to be able to interpret the QW conductance in terms of edge channels, we used a constant-voltage four-terminal diagonal measurement.\cite{Beenakker}

\section{RESULTS AND DISCUSSION \label {s3}}

\subsection{Dynamic nuclear polarization}

Figure \ref{f2}(a) shows the conductance difference $\Delta G$ (light traces and inset, left hand vertical axis) between magnetic-field up and down sweeps at a gate voltage $V_{\rm{SG}}=-0.8$ V. The field was swept at 10 T/h and the AC excitation was 100 $\mu$V at 77 Hz. This hysteresis appeared at the entrance and exit of the region where the filling factor (i. e. the ratio of electrons to magnetic flux quanta) in the QW $\nu_{\rm{w}}=1$, independent of gate voltage. We only present the data at $V_{\rm{SG}}=-0.8$ V, however, as the hysteresis at that gate voltage happened in the magnetic field region for which the Fermi energy in the 2DEG lied in a gap between Landau levels. We also performed conductance relaxation measurements by waiting for 5 minutes with no AC excitation voltage applied to the sample, then applying 100 $\mu$V AC and measuring the conductance for 120 seconds. The conductance relaxation $\Delta G$ at different magnetic fields can be seen in Fig. \ref{f2}(a) (grayscale plot, right hand vertical axis). $G$ shows the largest relaxation precisely just above and below $\nu_{\rm{w}}=1$ in the QW, coincidental with the conductance hysteresis. A vertical cut (dotted line) of Fig. \ref{f2}(a) can be seen in Fig. \ref{f2}(b). The relaxation time for the conductance is up to two minutes.

Both the hysteretic nature of the measurements and the long relaxation times suggest the involvement of nuclei in the effect. NMR signals for $^{75}$As (Figure \ref{f2}(c), signals vertically shifted for clarity) from 2 to 6.9 T prove this to be the case. NMR data were obtained by applying a continuous-wave RF voltage $V_{\rm{RF}}$ to the antenna overgate, sweeping the frequency $f_{\rm{RF}}$ upwards at 68 MHz/h and stepping the magnetic field in 0.1 T intervals while measuring the QW conductance. For increased clarity, Fig. \ref{f2}(d) shows the maximum positive gradient from the \textit{derivative} of the NMR signals in Fig. \ref{f2} (c). NMR signals are present just below 3 T, absent during the length of the $\nu_{\rm{w}}=1$ plateau and reappear strongly as $G$ leaves the plateau, vanishing [as do the relaxation traces in Fig. \ref{f2}(a)] near 6 T. The data points follow the linear relation $f_{\rm{RF}} = \gamma B$, where $\gamma$ is the gyromagnetic ratio of the nuclear species, 7.26 MHz/T for $^{75}$As. Similar signals (not shown) were observed for $^{69}$Ga and $^{71}$Ga.

\begin{figure}[h!]
\includegraphics[width=\columnwidth]{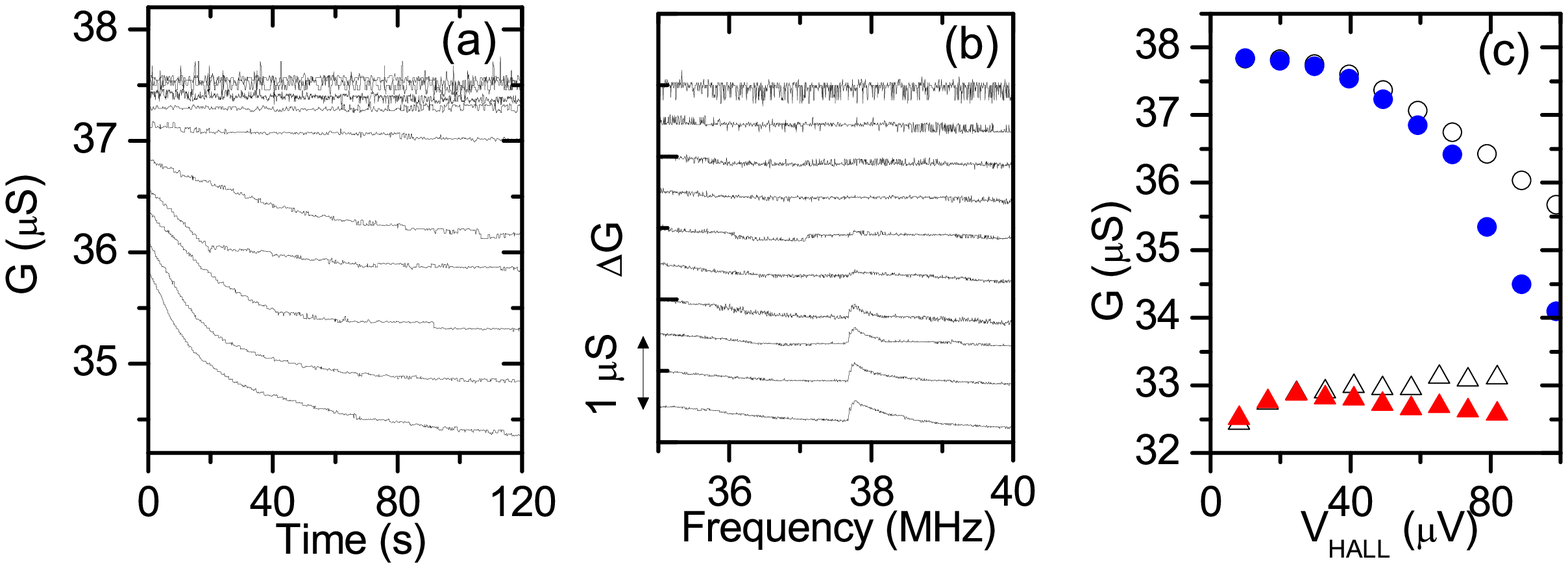}
\caption{(color online). (a) and (b) Onset of DNP with increasing excitation voltage as observed in conductance relaxation (a) and NMR ($^{75}$As) (b) measurements at $V_{\rm{SG}}=-0.8$ V and $B=5.2$ T. Excitation voltage varies in steps of 10 $\mu$V from 10 $\mu$V (uppermost curves) to 100 $\mu$V (lowermost curves). NMR lines have been shifted vertically for clarity; (c) Values of unpolarized (empty) and equilibrium (full) conductance values at 50 mK (circles) and 800 mK (triangles) (see text for explanation). \label{f3}}
\end{figure}

The DNP mechanism requires an electron to be scattered between energy states of opposite spins. The hyperfine interaction between electrons and nuclei allows for a spin change in the nuclear system to preserve total angular momentum, in a process that occurs locally where the electron spin-flip is induced. As more such scattering events occur, the DNP builds up and affects the energy structure of the electron system by creating an effective hyperfine field that adds to the external magnetic field, thus changing the electron Zeeman energy. In our experiment, the $\nu=2$ Landau subband is completely empty and electrons can scatter into it from the $\nu=1$ subband when a high enough voltage is applied accross the sample. The RF radiation at the NMR frequency randomizes the nuclei, eliminating the hyperfine field, and the corresponding change in Zeeman energy is observed in the conductance. At 5.2 T, where the strongest NMR signal is observed, we observed DNP thresholds for the relaxation [Figure \ref{f3}(a)] and NMR [Figure \ref{f3}(b)] measurements,\footnote{As in Fig. \ref{f2}(a), we waited for 5 minutes (to ensure zero nuclear polarization at $t=0$ s) before applying the excitation voltage in the relaxation measurements.} finding that nuclear polarization occurred for excitation voltages higher than about 40 $\mu$V. Note that these voltages are also the Hall voltages across the QW, since the 2DEG at 5.2 T has zero longitudinal resistivity. The threshold is smaller by roughly a factor of 3 from what would be expected according to the Zeeman-energy splitting, $\Delta E_{\rm{Z}}=|g|\mu_{B}B$, about 132 $\mu$eV at 5.2 T (using the bulk GaAs Land\'e \textit{g} factor $|g|=0.44$ and the Bohr magneton $\mu_{B}=58$ $\mu$eV/T). Figure \ref{f3}(c) shows the values of $G$ at $t=0$ s (which we will call \textit{unpolarized} value) and $t=120$ s (which we will call \textit{equilibrium} value) from Fig. \ref{f3}(a). We can see that DNP, although weaker, survives to high temperatures [triangles in Fig. \ref{f3}(c) at 800 mK] and presents an even lower threshold than at low temperatures.  Since no edge states are separately contacted in the 2DEG, the low threshold suggests that the DNP happens in or near the QW. In quantum dots in GaAs, the \textit{g} factor has been observed to be lower than the bulk.\cite{Baugh} Other studies in dots in InAs have also found the \textit{g} factor to be strongly dependent on dot size.\cite{Bjork} In GaAs quantum wires, however, enhancements of the \textit{g} factor due to exchange effects have been reported in the past.\cite{Thomas} Therefore, there is no clear understanding at present of the low DNP thresholds observed in our experiments. Other possibilities could be low-voltage breakdown of the quantum Hall effect\cite{BalevPRB} or electron correlation effects from edge states in narrow channels.\cite{BalevPRB2}

\begin{figure}
\includegraphics[width=\columnwidth]{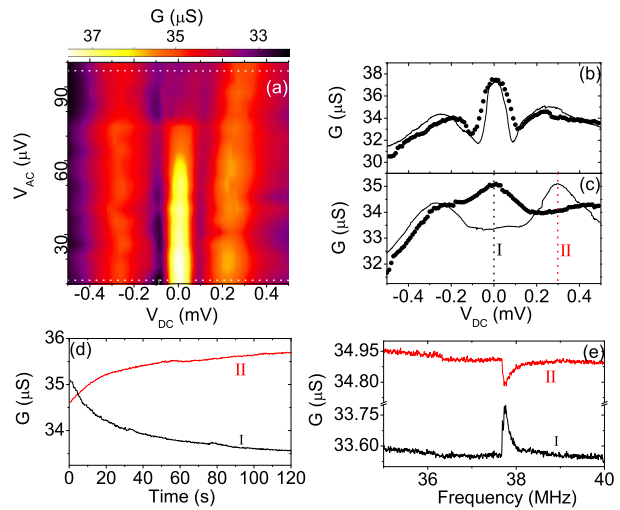}
\caption{(color online). (a) Conductance as a function of DC and AC excitation voltages; (b) and (c) Horizontal cuts of (a) at $V_{\rm{AC}}=10$ (b) and 95 (c) $\mu$V [dotted lines in (a)], showing the equilibrium (line) and unpolarized (dots) conductance as a function of DC Hall voltage; (d) and (e) Relaxation (d) and NMR (e) traces for $V_{\rm{DC}}$ values of 0 (I) and 0.3 (II) mV in (c). \label{f4}}
\end{figure}

All data presented in the remainder of this paper were taken at 5.2 T. In Figure \ref{f4} we can see the effect of applying a DC voltage along with the AC excitation. Data in Fig. \ref{f4}(a) were taken by sweeping the DC voltage upwards at 2 mV/h and increasing the AC excitation voltage by 5 $\mu$V between sweeps. Figures \ref{f4}(b) and \ref{f4}(c) show transverse cuts of Fig. \ref{f4}(a) at $V_{\rm{AC}}=10$ and $95 \mu$V, respectively. The difference between the dots (unpolarized) and lines (equilibrium) gives a measure of the amount of DNP. The $V_{\rm{DC}}$ threshold at low AC excitation is similar to the $V_{\rm{AC}}$ threshold without DC. As $V_{\rm{DC}}$ is raised near $\Delta E_{\rm{Z}}/e-V_{\rm{AC}}$, the $\nu_{\rm{w}}=2$ energy level starts being populated. As a consequence, the electron scattering rate from $\nu_{\rm{w}}=1$ into $\nu_{\rm{w}}=2$ decreases, whereas the electrons in $\nu_{\rm{w}}=2$ can now travel to the opposite side of the channel and scatter into $\nu_{\rm{w}}=1$ energy states. This results in a decrease of the observed DNP. As $V_{\rm{DC}}$ is increased and $\nu_{\rm{w}}=2$ filled, the DNP changes sign [Fig. \ref{f4}(c)]. The effect is absent for $V_{\rm{DC}}<0$, probably due to disorder, since we did observe it for both positive and negative $V_{\rm{DC}}$ in another sample. Polarization of both signs can be seen in Figs. \ref{f4}(d) and \ref{f4}(e), with curves I and II corresponding to 0 and 0.3 mV $V_{\rm{DC}}$, respectively [see Fig. \ref{f4}(c)].

\subsection{Pulsed NMR}

\begin{figure}
\includegraphics[width=\columnwidth]{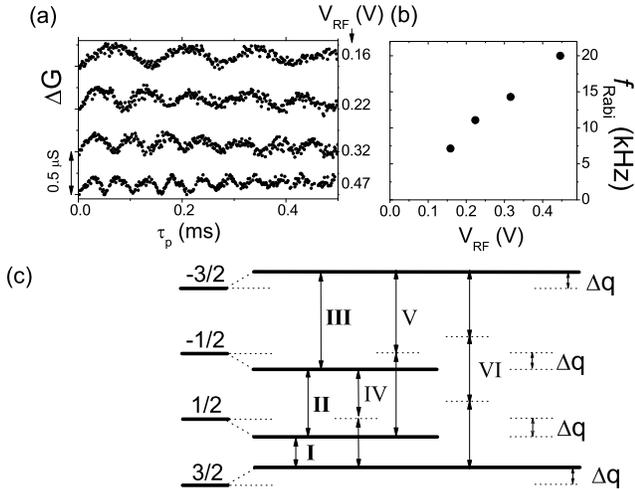}
\caption{(a) Rabi oscillations of $^{75}$As at 50 mK for $f_{\rm{RF}}=37.685$ MHz and $V_{\rm{RF}}= 0.16$, 0.22, 0.32 and 0.47 V ; (b) $f_{\rm{Rabi}}$ vs. $V_{\rm{RF}}$; (c) Energy levels diagram for a 3/2 spin with quadrupolar splitting, showing single- (I, II and III), two- (IV and V) and three- (VI) photon transitions. \label{f5}}
\end{figure}

Manipulation of $^{75}$As nuclei by pulsed NMR\cite{Machida, SanadaPRL, YusaNature} allows coherent control of nuclear states. A RF magnetic field $B_{\rm{RF}}$ is produced by applying an oscillating voltage $V_{\rm{RF}}$ to the antenna overgate for a pulse of duration $\tau_{\rm{p}}$.\footnote{A 100 $\mu$V AC excitation was kept on at all times during the pulse experiments to ensure maximum nuclear polarisation. Following each pulse, the conductance was allowed to relax for two minutes before applying the next pulse. The conductance was monitored at all times.} During the pulse, the nuclear quantum state precesses around the Bloch sphere at a frequency $f_{\rm{Rabi}}=\gamma B_{\rm{RF}}$. Figure \ref{f5}(a) shows coherent precession of $^{75}$As nuclear states, i.e. Rabi oscillations, at 50 mK for four different pulse amplitudes at a frequency $f_{\rm{RF}}=37.685$ MHz. This frequency is not exactly on resonance for the $^{75}$As spins and was chosen as a first rough approximation based on the NMR lines in Figs. \ref{f3}(b) and \ref{f4}(e). The dependence of $f_{\rm{Rabi}}$ with $V_{\rm{RF}}$ ($\propto B_{\rm{RF}}$) can be seen in Fig. \ref{f5}(b). From $f_{\rm{Rabi}}$ we estimate $B_{\rm{RF}}$ for each oscillation to be 0.98, 1.53, 1.97 and 2.75 mT.
 
\begin{figure}
\includegraphics[width=\columnwidth]{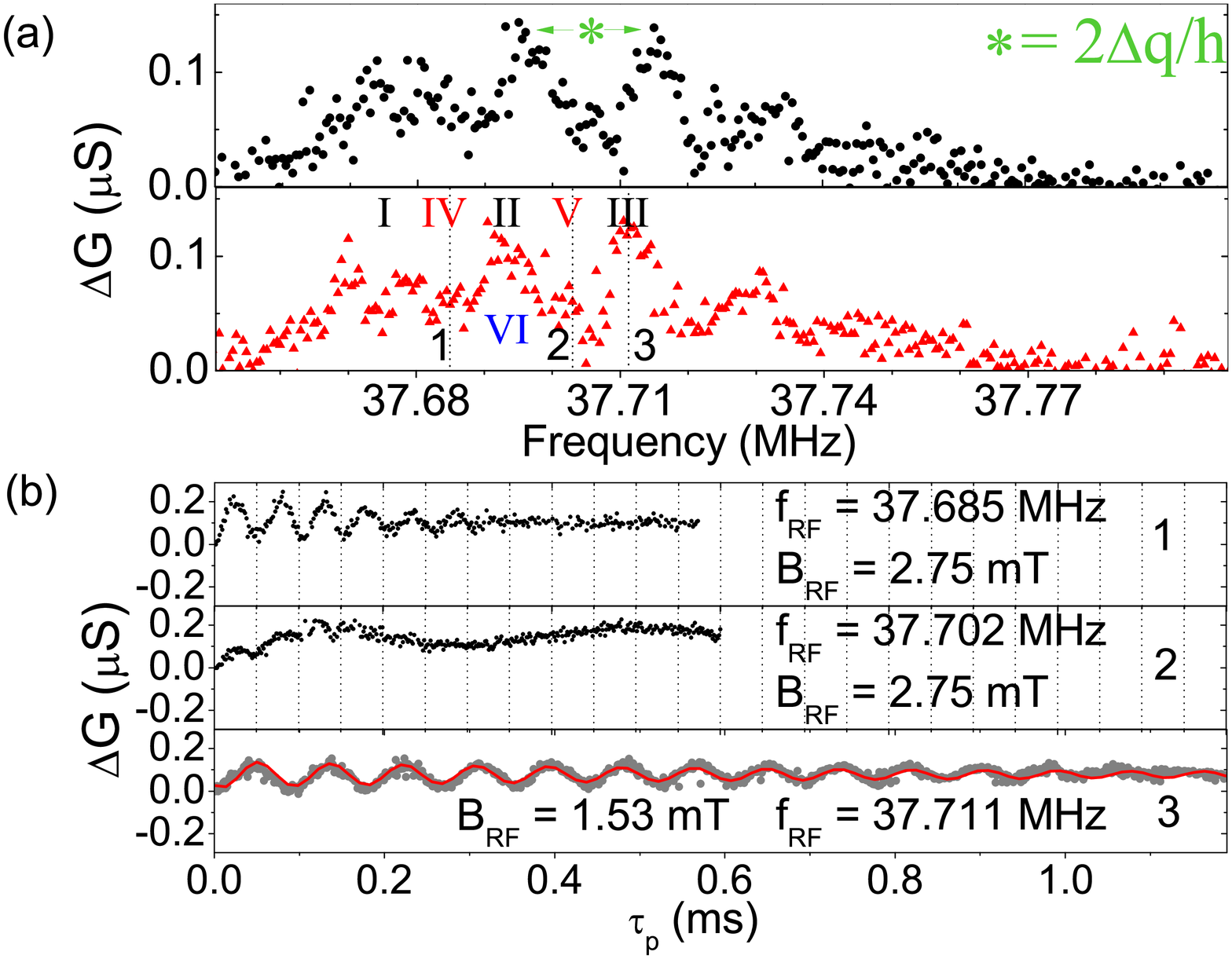}
\caption{(color online). (a) Frequency spectrum of $^{75}$As at $T=1.2$ K using 312 $\mu$s long pulses and $B_{\rm{RF}}=1.53$ mT with (triangles) and without (circles) electron-nuclear coupling, showing a 4 kHz Knight shift. Resonance frequencies for the different transitions in Fig. \ref{f5}(c) are indicated; (b) Rabi oscillations corresponding to the frequencies 1, 2 and 3 in (a).\label{f6}}
\end{figure}

We now show pulsed NMR data at $T$=1.2 K. By pinching-off the QW during the pulse so that the nuclei in the QW are no longer surrounded by conduction electrons during the nuclear precession we observe a Knight shift in the nuclear system.\cite{Slichter} Figure \ref{f6}(a) shows the NMR line with (lower, triangles) and without (upper, circles) electron-nuclear coupling. $132 \mu$s long pulses (equivalent to a $3\pi/2$ rotation), with $B_{\rm{RF}}=1.53$ mT, were applied while stepping $f_{\rm{RF}}$ by 0.5 kHz. The measured Knight shift was 4 kHz. If bulk were contributing to DNP, the peaks would broaden rather than simply shift. We only observe shifting, which supports the hypothesis of the DNP occurring in the QW or in its vicinity.

Quadrupolar interactions in the nuclear system due to the electric field gradient from the host lattice make various transitions at different energies possible, as depicted in Fig. \ref{f5}(c) [energies indicated in the spectrum in Fig. \ref{f6}(a)]. The $|\pm 3/2>$ nuclear states are shifted up by $\Delta_{\rm{q}}$ whereas the $|\pm 1/2>$ states are shifted down by the same amount. Transitions I, II and III are single-photon transitions, whereas transitions IV and V are two- and three-photon transitions, respectively (multiple-photon transitions require higher intensities of $B_{\rm{RF}}$ than single-photon transitions). Double and triple quantum coherence are, therefore, possible. From Fig. \ref{f6}(a) we measure a quadrupolar shift of $\Delta_{\rm{q}}/h=9$ kHz.

In Figure \ref{f6}(b) we show Rabi oscillations at 1.2 K for different $f_{\rm{RF}}$ and $B_{\rm{RF}}$. Of the two two-photon transitions, IV and V, only V can be excited at $B_{\rm{RF}}=2.75$ mT. The Rabi frequency for this transition is much lower than those of single-photon transitions. Transition II can be seen superimposed on transition V for short $\tau_{\rm{p}}$ in Fig. \ref{f6}(b,2). For short pulses both the resonant two-photon (V) and the off-resonant single-photon (II) transitions can be excited. As the pulse length increases, the single-photon transition II is no longer excited and disappears near $\tau_{\rm{p}}=0.2$ ms. Figure \ref{f6}(b,3) shows Rabi oscillations for $f_{\rm{RF}}=37.711$ MHz and $B_{\rm{RF}}=1.53$ mT. If we fit these data by $\Delta G \propto 1 - \cos(2\pi f_{\rm{Rabi}}\tau_{\rm{P}})\exp(-\tau_{\rm{P}}/T^{\rm{Rabi}}_{2})$ [solid line in Fig. \ref{f6}(b,3)], we find $f_{\rm{Rabi}}=11.71$ kHz and $T^{\rm{Rabi}}_{2}=0.82$ ms. Note that this are the only oscillations for which the pulse frequency was centered at a single-photon transition in our measurements. Both in Figure \ref{f5}(a) and in Figure \ref{f6}(b,1) and \ref{f6}(b,2) coherent evolution for single-photon transition seems to be shorter. The reason is that pulse bandwidth decreases with increasing pulse length and for longer pulses no single-photon transition is excited in the Rabi oscillations of Figure \ref{f5}(a) and Figure \ref{f6}(b,1) and \ref{f6}(b,2).

At 1.2 K, our measured $T^{\rm{Rabi}}_{2}=0.82$ ms is of the same order as that measured by Yusa \textit{et al}.\cite{YusaNature} at 50 mK, with the advantadge that our design does not require a back gate to tune electron density, and significantly longer than those from Takahashi \textit{et al}.\cite{Takahashi} and Sanada \textit{et al.}\cite{SanadaPRL} at 1.1 and 5.5 K, respectively. In contrast to Sanada \textit{et al}.,\cite{SanadaPRL} our measurements are controlled by all-electrical means and performed over a much smaller length scale.

Two main mechanisms could be responsible for the nuclear spin decoherence in our system: heteronuclear direct dipole coupling and electron-nuclear spin coupling. However, nuclear-nuclear direct coupling has a much smaller effect on nuclear decoherence than indirect electron-nuclear coupling (As-e-As and As-e-Ga).\cite{YusaEp} This is probably the reason why our observed Rabi oscillations are much longer than those from Takahashi \textit{et al}.,\cite{Takahashi} in which the nuclear system was in contact with a 2DEG. By contrast, in our QW the number of electrons interacting with each polarized nucleus is much smaller.

\section{CONCLUSIONS \label{s4}}

By using the breakdown of the QHE in a quantum wire, we have been able to dynamically polarize nuclear spins over a nanometer-scale. We found the thresholds for the onset of the breakdown of the QHE to be much lower than expected. This suggests electron correlation effects from edge states in the channel, although a detailed physical model is lacking at present. The dynamic nuclear polarization in this system could be used to create a local Overhauser field at relatively high temperatures with the ability of controlling its direction. 

In addition, we have achieved millisecond-long nuclear quantum coherence on a nanometer-scale by all-electrical means at 1.2 K. Single- and two-photon transitions were observed, and control over a four-level spin system suggests the possibility of performing two-qubit operations at temperatures around 1K. Our method could also possibly be extended to other materials that exhibit the QHE such as InSb or graphene. The sensitivity of our device, capable of detecting less than $\sim 10^8$ nuclear spins, is much higher than that of conventional NMR.\footnote{From the dimensions of our device ($700\times700$ nm$^{2}$ device area at definition, which is an upper limit, and a 2DEG thickness of $\sim$ 10 nm), we estimate the number of relevant nuclei in our measurements to be of the order of $\sim 10^8$, roughly one third of which are $^{75}$As nuclei.}

\begin{acknowledgments}
We thank N. Cooper, V. Tripathi, K. J. Thomas, F. Sfigakis, M. Kataoka, S. Chorley, K. das Gupta and T. Bergsten for insightful discussions and K. Cooper and A. Beckett for technical assistance. This work was supported by the EPSRC (UK).
\end{acknowledgments}

\bibliographystyle{apsrev}

\begin{thebibliography}{29}
\expandafter\ifx\csname natexlab\endcsname\relax\def\natexlab#1{#1}\fi
\expandafter\ifx\csname bibnamefont\endcsname\relax
  \def\bibnamefont#1{#1}\fi
\expandafter\ifx\csname bibfnamefont\endcsname\relax
  \def\bibfnamefont#1{#1}\fi
\expandafter\ifx\csname citenamefont\endcsname\relax
  \def\citenamefont#1{#1}\fi
\expandafter\ifx\csname url\endcsname\relax
  \def\url#1{\texttt{#1}}\fi
\expandafter\ifx\csname urlprefix\endcsname\relax\def\urlprefix{URL }\fi
\providecommand{\bibinfo}[2]{#2}
\providecommand{\eprint}[2][]{\url{#2}}

\bibitem[{\citenamefont{DiVincenzo}(1995)}]{DiVincenzo}
\bibinfo{author}{\bibfnamefont{D.~P.} \bibnamefont{DiVincenzo}},
  \bibinfo{journal}{Science} \textbf{\bibinfo{volume}{270}},
  \bibinfo{pages}{255} (\bibinfo{year}{1995}).

\bibitem[{\citenamefont{Kane}(1998)}]{KaneNature}
\bibinfo{author}{\bibfnamefont{B.~E.} \bibnamefont{Kane}},
  \bibinfo{journal}{Nature (London)} \textbf{\bibinfo{volume}{393}},
  \bibinfo{pages}{133} (\bibinfo{year}{1998}).

\bibitem[{\citenamefont{Leuenberger et~al.}(2002)\citenamefont{Leuenberger,
  Loss, Poggio, and Awschalom}}]{LeuenbergerLoss}
\bibinfo{author}{\bibfnamefont{M.~N.} \bibnamefont{Leuenberger}},
  \bibinfo{author}{\bibfnamefont{D.}~\bibnamefont{Loss}},
  \bibinfo{author}{\bibfnamefont{M.}~\bibnamefont{Poggio}}, \bibnamefont{and}
  \bibinfo{author}{\bibfnamefont{D.~D.} \bibnamefont{Awschalom}},
  \bibinfo{journal}{Phys. Rev. Lett.} \textbf{\bibinfo{volume}{89}},
  \bibinfo{pages}{207601} (\bibinfo{year}{2002}).

\bibitem[{\citenamefont{Taylor et~al.}(2003)\citenamefont{Taylor, Marcus, and
  Lukin}}]{TaylorMarcus}
\bibinfo{author}{\bibfnamefont{J.~M.} \bibnamefont{Taylor}},
  \bibinfo{author}{\bibfnamefont{C.~M.} \bibnamefont{Marcus}},
  \bibnamefont{and} \bibinfo{author}{\bibfnamefont{M.~D.} \bibnamefont{Lukin}},
  \bibinfo{journal}{Phys. Rev. Lett.} \textbf{\bibinfo{volume}{90}},
  \bibinfo{pages}{206803} (\bibinfo{year}{2003}).

\bibitem[{\citenamefont{Yusa et~al.}(2005)\citenamefont{Yusa, Muraki,
  Takashina, Hashimoto, and Hirayama}}]{YusaNature}
\bibinfo{author}{\bibfnamefont{G.}~\bibnamefont{Yusa}},
  \bibinfo{author}{\bibfnamefont{K.}~\bibnamefont{Muraki}},
  \bibinfo{author}{\bibfnamefont{K.}~\bibnamefont{Takashina}},
  \bibinfo{author}{\bibfnamefont{K.}~\bibnamefont{Hashimoto}},
  \bibnamefont{and} \bibinfo{author}{\bibfnamefont{Y.}~\bibnamefont{Hirayama}},
  \bibinfo{journal}{Nature (London)} \textbf{\bibinfo{volume}{434}},
  \bibinfo{pages}{1001} (\bibinfo{year}{2005}).

\bibitem[{\citenamefont{Deimling et~al.}(1980)\citenamefont{Deimling, Brunner,
  Dinse, and Hausser}}]{Deimling}
\bibinfo{author}{\bibfnamefont{M.}~\bibnamefont{Deimling}},
  \bibinfo{author}{\bibfnamefont{H.}~\bibnamefont{Brunner}},
  \bibinfo{author}{\bibfnamefont{K.~P.} \bibnamefont{Dinse}}, \bibnamefont{and}
  \bibinfo{author}{\bibfnamefont{K.~H.} \bibnamefont{Hausser}},
  \bibinfo{journal}{J. Magn. Reson.} \textbf{\bibinfo{volume}{39}},
  \bibinfo{pages}{185} (\bibinfo{year}{1980}).

\bibitem[{\citenamefont{Dirksen et~al.}(1989)\citenamefont{Dirksen, Henstra,
  and Wenckebach}}]{Dirksen}
\bibinfo{author}{\bibfnamefont{P.}~\bibnamefont{Dirksen}},
  \bibinfo{author}{\bibfnamefont{A.}~\bibnamefont{Henstra}}, \bibnamefont{and}
  \bibinfo{author}{\bibfnamefont{W.~T.} \bibnamefont{Wenckebach}},
  \bibinfo{journal}{J. Magn. Reson.} \textbf{\bibinfo{volume}{86}},
  \bibinfo{pages}{549} (\bibinfo{year}{1989}).

\bibitem[{\citenamefont{Tycko et~al.}(1995)\citenamefont{Tycko, Barrett,
  Dabbagh, Pfeiffer, and West}}]{Barrett}
\bibinfo{author}{\bibfnamefont{R.}~\bibnamefont{Tycko}},
  \bibinfo{author}{\bibfnamefont{S.~E.} \bibnamefont{Barrett}},
  \bibinfo{author}{\bibfnamefont{G.}~\bibnamefont{Dabbagh}},
  \bibinfo{author}{\bibfnamefont{L.~N.} \bibnamefont{Pfeiffer}},
  \bibnamefont{and} \bibinfo{author}{\bibfnamefont{K.~W.} \bibnamefont{West}},
  \bibinfo{journal}{Science} \textbf{\bibinfo{volume}{268}},
  \bibinfo{pages}{1460} (\bibinfo{year}{1995}).

\bibitem[{\citenamefont{Kikkawa and Awschalom}(2000)}]{Kikkawa}
\bibinfo{author}{\bibfnamefont{J.~M.} \bibnamefont{Kikkawa}} \bibnamefont{and}
  \bibinfo{author}{\bibfnamefont{D.~D.} \bibnamefont{Awschalom}},
  \bibinfo{journal}{Science} \textbf{\bibinfo{volume}{287}},
  \bibinfo{pages}{473} (\bibinfo{year}{2000}).

\bibitem[{\citenamefont{Sanada et~al.}(2006)\citenamefont{Sanada, Kondo,
  Matsuzaka, Morita, Hu, Ohno, and Ohno}}]{SanadaPRL}
\bibinfo{author}{\bibfnamefont{H.}~\bibnamefont{Sanada}},
  \bibinfo{author}{\bibfnamefont{Y.}~\bibnamefont{Kondo}},
  \bibinfo{author}{\bibfnamefont{S.}~\bibnamefont{Matsuzaka}},
  \bibinfo{author}{\bibfnamefont{K.}~\bibnamefont{Morita}},
  \bibinfo{author}{\bibfnamefont{C.~Y.} \bibnamefont{Hu}},
  \bibinfo{author}{\bibfnamefont{Y.}~\bibnamefont{Ohno}}, \bibnamefont{and}
  \bibinfo{author}{\bibfnamefont{H.}~\bibnamefont{Ohno}},
  \bibinfo{journal}{Phys. Rev. Lett.} \textbf{\bibinfo{volume}{96}},
  \bibinfo{pages}{067602} (\bibinfo{year}{2006}).

\bibitem[{\citenamefont{Kondo et~al.}(2008)\citenamefont{Kondo, Ono, Matsuzaka,
  Morita, Sanada, Ohno, and Ohno}}]{Kondo}
\bibinfo{author}{\bibfnamefont{Y.}~\bibnamefont{Kondo}},
  \bibinfo{author}{\bibfnamefont{M.}~\bibnamefont{Ono}},
  \bibinfo{author}{\bibfnamefont{S.}~\bibnamefont{Matsuzaka}},
  \bibinfo{author}{\bibfnamefont{K.}~\bibnamefont{Morita}},
  \bibinfo{author}{\bibfnamefont{H.}~\bibnamefont{Sanada}},
  \bibinfo{author}{\bibfnamefont{Y.}~\bibnamefont{Ohno}}, \bibnamefont{and}
  \bibinfo{author}{\bibfnamefont{H.}~\bibnamefont{Ohno}},
  \bibinfo{journal}{Phys. Rev. Lett.} \textbf{\bibinfo{volume}{101}},
  \bibinfo{pages}{207601} (\bibinfo{year}{2008}).

\bibitem[{\citenamefont{Machida et~al.}(2003)\citenamefont{Machida, Yamazaki,
  Ikushima, and Komiyama}}]{Machida}
\bibinfo{author}{\bibfnamefont{T.}~\bibnamefont{Machida}},
  \bibinfo{author}{\bibfnamefont{T.}~\bibnamefont{Yamazaki}},
  \bibinfo{author}{\bibfnamefont{K.}~\bibnamefont{Ikushima}}, \bibnamefont{and}
  \bibinfo{author}{\bibfnamefont{S.}~\bibnamefont{Komiyama}},
  \bibinfo{journal}{Appl. Phys. Lett.} \textbf{\bibinfo{volume}{82}},
  \bibinfo{pages}{409} (\bibinfo{year}{2003}).

\bibitem[{\citenamefont{Dixon et~al.}(1997)\citenamefont{Dixon, Wald, McEuen,
  and Melloch}}]{Dixon}
\bibinfo{author}{\bibfnamefont{D.~C.} \bibnamefont{Dixon}},
  \bibinfo{author}{\bibfnamefont{K.~R.} \bibnamefont{Wald}},
  \bibinfo{author}{\bibfnamefont{P.~L.} \bibnamefont{McEuen}},
  \bibnamefont{and} \bibinfo{author}{\bibfnamefont{M.~R.}
  \bibnamefont{Melloch}}, \bibinfo{journal}{Phys. Rev. B}
  \textbf{\bibinfo{volume}{56}}, \bibinfo{pages}{4743} (\bibinfo{year}{1997}).

\bibitem[{\citenamefont{Bowers et~al.}(2006)\citenamefont{Bowers, Caldwell,
  Gusev, Kovalev, Olshanetsky, Reno, Simmons, and Vitkalov}}]{Bowers}
\bibinfo{author}{\bibfnamefont{C.~R.} \bibnamefont{Bowers}},
  \bibinfo{author}{\bibfnamefont{J.~D.} \bibnamefont{Caldwell}},
  \bibinfo{author}{\bibfnamefont{G.}~\bibnamefont{Gusev}},
  \bibinfo{author}{\bibfnamefont{A.~E.} \bibnamefont{Kovalev}},
  \bibinfo{author}{\bibfnamefont{E.}~\bibnamefont{Olshanetsky}},
  \bibinfo{author}{\bibfnamefont{J.~L.} \bibnamefont{Reno}},
  \bibinfo{author}{\bibfnamefont{J.~A.} \bibnamefont{Simmons}},
  \bibnamefont{and} \bibinfo{author}{\bibfnamefont{S.~A.}
  \bibnamefont{Vitkalov}}, \bibinfo{journal}{Solid State Nucl. Mag.}
  \textbf{\bibinfo{volume}{29}}, \bibinfo{pages}{52} (\bibinfo{year}{2006}).

\bibitem[{\citenamefont{Olshanetsky et~al.}(2006)\citenamefont{Olshanetsky,
  Caldwell, Kovalev, Bowers, Simmons, and Reno}}]{Olshanetsky}
\bibinfo{author}{\bibfnamefont{E.}~\bibnamefont{Olshanetsky}},
  \bibinfo{author}{\bibfnamefont{J.~D.} \bibnamefont{Caldwell}},
  \bibinfo{author}{\bibfnamefont{A.~E.} \bibnamefont{Kovalev}},
  \bibinfo{author}{\bibfnamefont{C.~R.} \bibnamefont{Bowers}},
  \bibinfo{author}{\bibfnamefont{J.~A.} \bibnamefont{Simmons}},
  \bibnamefont{and} \bibinfo{author}{\bibfnamefont{J.~L.} \bibnamefont{Reno}},
  \bibinfo{journal}{Physica B} \textbf{\bibinfo{volume}{373}},
  \bibinfo{pages}{182} (\bibinfo{year}{2006}).

\bibitem[{\citenamefont{Smet et~al.}(2002)\citenamefont{Smet, Deutschmann,
  Ertl, Wegscheider, Abstreiter, and von Klitzing}}]{Smet}
\bibinfo{author}{\bibfnamefont{J.~H.} \bibnamefont{Smet}},
  \bibinfo{author}{\bibfnamefont{R.~A.} \bibnamefont{Deutschmann}},
  \bibinfo{author}{\bibfnamefont{F.}~\bibnamefont{Ertl}},
  \bibinfo{author}{\bibfnamefont{W.}~\bibnamefont{Wegscheider}},
  \bibinfo{author}{\bibfnamefont{G.}~\bibnamefont{Abstreiter}},
  \bibnamefont{and} \bibinfo{author}{\bibfnamefont{K.}~\bibnamefont{von
  Klitzing}}, \bibinfo{journal}{Nature (London)}
  \textbf{\bibinfo{volume}{415}}, \bibinfo{pages}{281} (\bibinfo{year}{2002}).

\bibitem[{\citenamefont{Kawamura et~al.}(2007)\citenamefont{Kawamura,
  Takahashi, Sugihara, Masubuchi, Hamaya, and Machida}}]{Kawamura}
\bibinfo{author}{\bibfnamefont{M.}~\bibnamefont{Kawamura}},
  \bibinfo{author}{\bibfnamefont{H.}~\bibnamefont{Takahashi}},
  \bibinfo{author}{\bibfnamefont{K.}~\bibnamefont{Sugihara}},
  \bibinfo{author}{\bibfnamefont{S.}~\bibnamefont{Masubuchi}},
  \bibinfo{author}{\bibfnamefont{K.}~\bibnamefont{Hamaya}}, \bibnamefont{and}
  \bibinfo{author}{\bibfnamefont{T.}~\bibnamefont{Machida}},
  \bibinfo{journal}{Appl. Phys. Lett.} \textbf{\bibinfo{volume}{90}},
  \bibinfo{pages}{022102} (\bibinfo{year}{2007}).

\bibitem[{\citenamefont{Takahashi et~al.}(2007)\citenamefont{Takahashi,
  Kawamura, Masubuchi, Hamaya, and Machida}}]{Takahashi}
\bibinfo{author}{\bibfnamefont{H.}~\bibnamefont{Takahashi}},
  \bibinfo{author}{\bibfnamefont{M.}~\bibnamefont{Kawamura}},
  \bibinfo{author}{\bibfnamefont{S.}~\bibnamefont{Masubuchi}},
  \bibinfo{author}{\bibfnamefont{K.}~\bibnamefont{Hamaya}}, \bibnamefont{and}
  \bibinfo{author}{\bibfnamefont{T.}~\bibnamefont{Machida}},
  \bibinfo{journal}{Appl. Phys. Lett.} \textbf{\bibinfo{volume}{91}},
  \bibinfo{pages}{092120} (\bibinfo{year}{2007}).

\bibitem[{\citenamefont{Ladd et~al.}(2005)\citenamefont{Ladd, Maryenko,
  Yamamoto, Abe, and Itoh}}]{LaddPRB}
\bibinfo{author}{\bibfnamefont{T.~D.} \bibnamefont{Ladd}},
  \bibinfo{author}{\bibfnamefont{D.}~\bibnamefont{Maryenko}},
  \bibinfo{author}{\bibfnamefont{Y.}~\bibnamefont{Yamamoto}},
  \bibinfo{author}{\bibfnamefont{E.}~\bibnamefont{Abe}}, \bibnamefont{and}
  \bibinfo{author}{\bibfnamefont{K.~M.} \bibnamefont{Itoh}},
  \bibinfo{journal}{Phys. Rev. B} \textbf{\bibinfo{volume}{71}},
  \bibinfo{pages}{014401} (\bibinfo{year}{2005}).

\bibitem[{\citenamefont{van Wees et~al.}(1988)\citenamefont{van Wees, van
  Houten, Beenakker, Williamson, Kouwenhoven, van~der Marel, and
  Foxon}}]{vanWees}
\bibinfo{author}{\bibfnamefont{B.~J.} \bibnamefont{van Wees}},
  \bibinfo{author}{\bibfnamefont{H.}~\bibnamefont{van Houten}},
  \bibinfo{author}{\bibfnamefont{C.~W.~J.} \bibnamefont{Beenakker}},
  \bibinfo{author}{\bibfnamefont{J.~G.} \bibnamefont{Williamson}},
  \bibinfo{author}{\bibfnamefont{L.~P.} \bibnamefont{Kouwenhoven}},
  \bibinfo{author}{\bibfnamefont{D.}~\bibnamefont{van~der Marel}},
  \bibnamefont{and} \bibinfo{author}{\bibfnamefont{C.~T.} \bibnamefont{Foxon}},
  \bibinfo{journal}{Phys. Rev. Lett.} \textbf{\bibinfo{volume}{60}},
  \bibinfo{pages}{848} (\bibinfo{year}{1988}).

\bibitem[{\citenamefont{Wharam et~al.}(1988)\citenamefont{Wharam, Thornton,
  Newbury, Pepper, Ahmed, Frost, Hasko, Peacock, Ritchie, and Jones}}]{Wharam}
\bibinfo{author}{\bibfnamefont{D.~A.} \bibnamefont{Wharam}},
  \bibinfo{author}{\bibfnamefont{T.~J.} \bibnamefont{Thornton}},
  \bibinfo{author}{\bibfnamefont{R.}~\bibnamefont{Newbury}},
  \bibinfo{author}{\bibfnamefont{M.}~\bibnamefont{Pepper}},
  \bibinfo{author}{\bibfnamefont{H.}~\bibnamefont{Ahmed}},
  \bibinfo{author}{\bibfnamefont{J.~E.~F.} \bibnamefont{Frost}},
  \bibinfo{author}{\bibfnamefont{D.~G.} \bibnamefont{Hasko}},
  \bibinfo{author}{\bibfnamefont{D.~C.} \bibnamefont{Peacock}},
  \bibinfo{author}{\bibfnamefont{D.~A.} \bibnamefont{Ritchie}},
  \bibnamefont{and} \bibinfo{author}{\bibfnamefont{G.~A.~C.}
  \bibnamefont{Jones}}, \bibinfo{journal}{J. Phys. C}
  \textbf{\bibinfo{volume}{21}}, \bibinfo{pages}{L209} (\bibinfo{year}{1988}).

\bibitem[{\citenamefont{Beenakker and van Houten}(1991)}]{Beenakker}
\bibinfo{author}{\bibfnamefont{C.~W.~J.} \bibnamefont{Beenakker}}
  \bibnamefont{and} \bibinfo{author}{\bibfnamefont{H.}~\bibnamefont{van
  Houten}}, \emph{\bibinfo{title}{Quantum Transport in Semiconductor
  Nanostructures}}, vol.~\bibinfo{volume}{44} (\bibinfo{publisher}{Solid State
  Phys.}, \bibinfo{year}{1991}).

\bibitem[{\citenamefont{Baugh et~al.}(2007)\citenamefont{Baugh, Kitamura, Ono,
  and Tarucha}}]{Baugh}
\bibinfo{author}{\bibfnamefont{J.}~\bibnamefont{Baugh}},
  \bibinfo{author}{\bibfnamefont{Y.}~\bibnamefont{Kitamura}},
  \bibinfo{author}{\bibfnamefont{K.}~\bibnamefont{Ono}}, \bibnamefont{and}
  \bibinfo{author}{\bibfnamefont{S.}~\bibnamefont{Tarucha}},
  \bibinfo{journal}{Phys. Rev. Lett.} \textbf{\bibinfo{volume}{99}},
  \bibinfo{pages}{096804} (\bibinfo{year}{2007}).

\bibitem[{\citenamefont{Bj\"{o}rk et~al.}(2005)\citenamefont{Bj\"{o}rk, Fuhrer,
  Hansen, Larsson, Fr\"{o}berg, and Samuelson}}]{Bjork}
\bibinfo{author}{\bibfnamefont{M.~T.} \bibnamefont{Bj\"{o}rk}},
  \bibinfo{author}{\bibfnamefont{A.}~\bibnamefont{Fuhrer}},
  \bibinfo{author}{\bibfnamefont{A.~E.} \bibnamefont{Hansen}},
  \bibinfo{author}{\bibfnamefont{M.~W.} \bibnamefont{Larsson}},
  \bibinfo{author}{\bibfnamefont{L.~E.} \bibnamefont{Fr\"{o}berg}},
  \bibnamefont{and}
  \bibinfo{author}{\bibfnamefont{L.}~\bibnamefont{Samuelson}},
  \bibinfo{journal}{Phys. Rev. B} \textbf{\bibinfo{volume}{72}},
  \bibinfo{pages}{201307(R)} (\bibinfo{year}{2005}).

\bibitem[{\citenamefont{Thomas et~al.}(1998)\citenamefont{Thomas, Nicholls,
  Appleyard, Simmons, Pepper, Mace, Tribe, and Ritchie}}]{Thomas}
\bibinfo{author}{\bibfnamefont{K.~J.} \bibnamefont{Thomas}},
  \bibinfo{author}{\bibfnamefont{J.~T.} \bibnamefont{Nicholls}},
  \bibinfo{author}{\bibfnamefont{N.~J.} \bibnamefont{Appleyard}},
  \bibinfo{author}{\bibfnamefont{M.~Y.} \bibnamefont{Simmons}},
  \bibinfo{author}{\bibfnamefont{M.}~\bibnamefont{Pepper}},
  \bibinfo{author}{\bibfnamefont{D.~R.} \bibnamefont{Mace}},
  \bibinfo{author}{\bibfnamefont{W.~R.} \bibnamefont{Tribe}}, \bibnamefont{and}
  \bibinfo{author}{\bibfnamefont{D.~A.} \bibnamefont{Ritchie}},
  \bibinfo{journal}{Phys. Rev. B} \textbf{\bibinfo{volume}{58}},
  \bibinfo{pages}{4846} (\bibinfo{year}{1998}).

\bibitem[{\citenamefont{Balev et~al.}(1994)\citenamefont{Balev, Vasilopoulos,
  and Mozdor}}]{BalevPRB}
\bibinfo{author}{\bibfnamefont{O.~G.} \bibnamefont{Balev}},
  \bibinfo{author}{\bibfnamefont{P.}~\bibnamefont{Vasilopoulos}},
  \bibnamefont{and} \bibinfo{author}{\bibfnamefont{E.~V.}
  \bibnamefont{Mozdor}}, \bibinfo{journal}{Phys. Rev. B}
  \textbf{\bibinfo{volume}{50}}, \bibinfo{pages}{8706} (\bibinfo{year}{1994}).

\bibitem[{\citenamefont{Balev and Studart}(2001)}]{BalevPRB2}
\bibinfo{author}{\bibfnamefont{O.~G.} \bibnamefont{Balev}} \bibnamefont{and}
  \bibinfo{author}{\bibfnamefont{N.}~\bibnamefont{Studart}},
  \bibinfo{journal}{Phys. Rev. B} \textbf{\bibinfo{volume}{64}},
  \bibinfo{pages}{115309} (\bibinfo{year}{2001}).

\bibitem[{\citenamefont{Slichter}(1990)}]{Slichter}
\bibinfo{author}{\bibfnamefont{C.~P.} \bibnamefont{Slichter}},
  \emph{\bibinfo{title}{Principles of Magnetic Resonance}}
  (\bibinfo{publisher}{Springer}, \bibinfo{address}{Berlin},
  \bibinfo{year}{1990}).

\bibitem[{\citenamefont{Yusa et~al.}(2006)\citenamefont{Yusa, Kumada, Muraki,
  and Hirayama}}]{YusaEp}
\bibinfo{author}{\bibfnamefont{G.}~\bibnamefont{Yusa}},
  \bibinfo{author}{\bibfnamefont{N.}~\bibnamefont{Kumada}},
  \bibinfo{author}{\bibfnamefont{K.}~\bibnamefont{Muraki}}, \bibnamefont{and}
  \bibinfo{author}{\bibfnamefont{Y.}~\bibnamefont{Hirayama}},
  \bibinfo{journal}{arXiv:cond-mat/0510310}  (\bibinfo{year}{2006}).

\end{thebibliography}

\end{document}